\documentclass[pra,nofootinbib,twocolumn,amsmath,amssymb,superscriptaddress]{revtex4-2}
\usepackage{graphicx,amsmath,relsize,epstopdf,color,mathtools,bm,newtxtext,newtxmath,braket,rotating}
\usepackage[hyphenbreaks]{breakurl}
\usepackage[colorlinks=true,linkcolor=blue,citecolor=blue,urlcolor =blue]{hyperref}
\usepackage[normalem]{ulem}

\usepackage{booktabs}
\usepackage[table,xcdraw]{xcolor}

\newcommand{\IM}{\mathop{\mathrm{Im}} \nolimits}
\newcommand{\Tr}{\mathop{\mathrm{Tr}} \nolimits}

\usepackage{soul,xcolor}

\usepackage{dsfont}

\begin{document}
	
\title{Sensing rotations with multiplane light conversion}
	
\author{M. Eriksson}
\affiliation{Physics Unit, Photonics Laboratory, Tampere University, 33720 Tampere, Finland}
	
\author{A. Z. Goldberg}
\affiliation{National Research Council of Canada, 100 Sussex Drive, Ottawa, Ontario K1N 5A2, Canada}
\affiliation{Department of Physics, University of Ottawa, 25 Templeton Street, Ottawa, Ontario, K1N 6N5 Canada}

\author{M. Hiekkam\"aki}
\affiliation{Physics Unit, Photonics Laboratory, Tampere University, 33720 Tampere, Finland}
	
\author{F.~Bouchard}
\affiliation{National Research Council of Canada, 100 Sussex Drive, Ottawa, Ontario K1N 5A2, Canada}

\author{G. Leuchs}
\affiliation{Max-Planck-Institut für die Physik des Lichts, 91058 Erlangen, Germany}
\affiliation{Department of Physics, Friedrich-Alexander-Universit\"at Erlangen-N\"urnberg, 91058 Erlangen, Germany}

\author{R. Fickler}
\affiliation{Physics Unit, Photonics Laboratory, Tampere University, 33720 Tampere, Finland}

\author{L. L. S\'anchez-Soto}
\affiliation{Max-Planck-Institut für die Physik des Lichts, 91058 Erlangen, Germany}
\affiliation{Departamento de \'Optica, Facultad de F\'{\i}sica, Universidad Complutense, 28040 Madrid, Spain}

\begin{abstract}
We report an experiment estimating the three parameters of a general rotation. The scheme uses quantum states attaining the ultimate precision dictated by the quantum Cram\'er-Rao bound. We realize the states experimentally using the orbital angular momentum of light and implement the rotations with a multiplane light conversion setup, which allows one to perform arbitrary unitary transformations on a finite set of spatial modes.  The observed performance suggests a range of potential applications in the next generation of rotation sensors.
\end{abstract}
	
\maketitle

\section{Introduction} 

Rotation sensors are indispensable elements for numerous applications. Examples include inertial navigation~\cite{Grewal:2013aa,Lawrence:1998aa}, geophysical studies~\cite{Stedman:1997aa}, and tests of general relativity~\cite{Leuchs:1986aa,Cerdonio:1988aa,Ciufolini:1998aa}, and significant technological progress is opening novel potential uses. As diverse as the applications are, the repertoire of available sensors has continued to grow: from small MEMS gyros~\cite{Zhanshe:2015aa}, over fiber-optic gyros~\cite{Lefere:2014aa} and electrochemical devices~\cite{Agafonov:2014aa,Leugoud:2012aa,Cusano:2016aa}, to high-resolution ring lasers~\cite{Chow:1985aa,Anderson:1994aa} and matter-wave interferometers~\cite{Gustavson:2000aa,Durfee:2006aa,Savoie:2018aa}.

These technological advances are boosting the performance to levels where quantum effects come into play~\cite{Degen:2017aa}. Therefore, it seems pertinent to analyze the ultimate limits of rotation sensors from a quantum perspective. 

The problem of determining all three parameters defining a rotation constitutes a paradigmatic example of a multiparameter estimation. This  encapsulates the confluence between measuring incompatible observables and achieving the ultimate precision limits.  Quantum metrology promises that certain probe states and measurement procedures can dramatically outperform standard (classical) protocols to simultaneously estimate  multiple parameters with the ultimate precision~\cite{Szczykulska:2016aa,Sidhu:2020aa,Albarelli:2020aa,Polino:2020aa,Rafal:2020aa}.

The optimal probe states for sensing arbitrary rotations are known as Kings of Quantumness~\cite{Bjork:2015aa,Bjork:2015ab} (initially dubbed anticoherent states~\cite{Zimba:2006fk}). These probes have the remarkable property that their low-order moments of important observables are \textit{unchanged} via rotations, instead imprinting rotation information in higher-order moments. As well, through the Majorana representation~\cite{Majorana:1932ul}, they exhibit highly symmetric geometrical structures on the Poincar\'e (or Bloch) sphere, which can be used to intuit their metrological properties~\cite{Hoz:2013om,Hoz:2014kq,Goldberg:2020ac}. These probe states have previously been generated with orbital angular momentum (OAM) carrying light modes to demonstrate single-parameter estimation~\cite{Bouchard:2017aa} and, recently, also in intrinsic polarization degrees of freedom~\cite{Ferretti:2022aa}. 

Given an optimal probe state subject to an unknown rotation, what measurement scheme best reveals the rotation parameters? Although there might be an ideal positive operator-valued measure (POVM) that may never be easily realized in a realistic system, we can aptly ask where it can be approximated with straightforward measurements. The answer is positive: just like a globe can be oriented by finding the locations of London and Tokyo, so, too, can a quantum state be oriented by measuring its projections onto a small number of axes pointed at various directions on the surface of the sphere. This scheme was outlined in Ref.~\cite{Goldberg:2021aa} and is now demonstrated for the first time.

We generate our ideal probe states in the OAM basis by using spatial light modulators (SLMs). This basis comprises a high-dimensional state space. These states then undergo rotations by passing through a multiplane light converter (MPLC)~\cite{Morizur:2010aa,Boucher:2021aa}, which is capable of enacting arbitrary {linear} transformations using a series of phase modulations with free-space propagation between each of the planes. The rotated states are then projected onto coherent states oriented along various axes; from these data, we can reconstruct all of the rotation parameters.

This paper is organized as follows. In Sec.~\ref{Sec:prel}, we briefly review some basic properties of rotations and their effects. In Sec.~\ref{Sec:est}, we discuss the ultimate limits in rotation sensing and derive the optimal states for that task.  In Sec.~\ref{Sec:exp} we present the details of our experimental setup, while, in Sec.~\ref{Sec:res}, we analyze the obtained results. Finally, our conclusions are summarized in Sec.~\ref{Sec:concl}.

\section{Preliminaries about rotations}
\label{Sec:prel}

In general, a rotation is characterized by three parameters~\cite{Grafarend:2011aa}: either the two angular coordinates of the rotation axis and the angle rotated around that axis, or the Euler angles. We follow  the former option throughout and consider a rotation of angle $\omega$ and rotation axis $\mathbf{u}(\Theta,\Phi)=\begin{pmatrix}\sin\Theta \cos\Phi & \sin\Theta\sin\Phi &\cos\Theta\end{pmatrix}^\top$, where the superscript $\top$ denotes the transpose. We will use the compact notation $\bm{\Omega} (\omega,\mathbf{u}) = (\omega,\Theta, \Phi)$ to denote these angles. It is well known that the action of this rotation in Hilbert space is represented by~\cite{Cornwell:1984aa} 
\begin{equation}
R ( \bm{\Omega}) = e^{ i \omega\mathbf{J} \cdot \mathbf{u}} \, , 
\end{equation}
where we have used the standard angular momentum notation $\mathbf{J}$  for the generators, which  satisfy the commutation relations of the Lie algebra $\mathfrak{su}(2)$:  $[J_{x}, J_{y}] = i  J_{z}$ and circular permutations (with $\hbar = 1$).

We consider the $(2J+1)-$dimensional space $\mathcal{H}_{J}$, spanned  by the states $\{ |J m\rangle \}$, with $m = -J, \ldots, +J$. This is the Hilbert space of spin-$J$ particles, but also describes the case of $2J$ qubits. Indeed, via the Jordan-Schwinger representation~\cite{Jordan:1935aa,Schwinger:1965kx}, which represents the algebra $\mathfrak{su}(2)$ in terms of bosonic amplitudes, the space $\mathcal{H}_{J}$ also encompasses many different instances of two-mode problems, such as, e.g., polarization, strongly correlated systems, and Bose-Einstein condensates ~\cite{Chaturvedi:2006vn} with fixed total numbers of excitations. Actually, one can consider the spin $J$ as a proxy for the input resources required in a metrological setting and inspect the precision of various estimates in terms of $J$. In what follows, we  assume that we work in $\mathcal{H}_{J}$. This restriction is reasonable since maximal precision will be obtained by concentrating all of the resources into a single subspace corresponding to the average total number of particles.

The notion of Majorana constellation{s}~\cite{Majorana:1932ul} will prove to be extremely convenient for our purposes. In this representation, a pure state corresponds to a configuration of points on the Bloch sphere, a picture that makes a high{-}dimensional Hilbert space easier to comprehend. The idea can be presented in a variety of ways~\cite{Bacry:2004aa,Bengtsson:2017aa}, but the most direct one is, perhaps, by first recalling that SU(2)- or Bloch-coherent states can be defined as~\cite{Perelomov:1986ly,Gazeau:2009aa} 
\begin{equation}
|z \rangle \equiv |\mathbf{n} \rangle = \frac{1}{(1 + | z |^{2})^{J}}  \exp(z J_{-}) |JJ \rangle \,, 
\end{equation}
where $J_{\pm} = J_{x} \pm i J_{y}$ are ladder operators and $z = \exp(i \phi) \, \tan (\theta/{2}) $, which is an inverse stereographical mapping from $z \in \mathbb{C}$ to the point of spherical coordinates $(\theta, \phi)$ fixing  the unit vector $\mathbf{n}$. Coherent states are precisely eigenstates of the operator $\mathbf{J} \cdot \mathbf{n}$ and they constitute an overcomplete basis. So, every pure state $|\psi \rangle \in \mathcal{H}_{J}$  can be expanded in that basis as 
\begin{equation}
  \label{eq:MajPol}
  \psi (z) = \langle z | \psi \rangle =  \frac{1}{(1 + | z |^{2})^{J}}
  \sum_{m=-J}^{J} {2J \choose J+m}^{1/2} \;
  \psi_{m} \, z^{J+m} \, ,
\end{equation}
where $\psi_{m} = \langle J m |\psi \rangle$ are the amplitudes of the state in the angular momentum  basis. Since this is a polynomial, $| \psi \rangle$ is determined by the set $\{ z_{i} \}$ of the $2J$ complex zeros of $\psi (z)$.  A nice geometrical representation of $| \psi \rangle$ by $2J$ points on the unit sphere (often called the Majorana constellation) is obtained by an inverse stereographic map $\{ z_{i}\} \mapsto \{ \theta_{i}, \phi_{i} \}$.

Let us examine a few examples to illustrate how this representation works in practice. The first one is that of SU(2) or Bloch coherent states $\ket{\mathbf{n}_{0}}$, for which the constellation collapses in this case to a single point diametrically opposed to $ \mathbf{n}_{0}$. 

For the angular momentum basis, $\ket{Jm}$ can be easily inferred from the polynomial so they consist of $J\pm m$ stars at the north and south poles, respectively. 

Another relevant set of states are the NOON states~\cite{Dowling:2008aa} 
\begin{equation}
 |\mathrm{NOON} \rangle =  \frac{1}{\sqrt{2}}
  {(|JJ\rangle - |J\, -\!J\rangle)}  ,
  \label{Eq: NOON}
\end{equation}
for which the Majorana constellations have $2J$ stars placed around the equator of the Bloch sphere with equal angular separation between each star.

Since the most classical states  (i.e., coherent states) have the most concentrated constellation, one might intuitively think that the most quantum states have their $2J$ stars distributed most symmetrically on the unit sphere, and this is the case. This constitutes the realm of the Kings of Quantumness~\cite{Bjork:2015aa,Bjork:2015ab}. In a sense they are the opposite of Bloch coherent states, as they \emph{point nowhere}; i.e., the average angular momentum vanishes and the fluctuations up to given order $M$ are isotropic~\cite{Hoz:2013om,Hoz:2014kq,Goldberg:2020ac}. Their symmetrical Majorana constellations herald their isotropic angular momentum properties and {give} an intuitive picture that these states are the most sensitive for rotation measurements.

\section{Estimating rotation parameters}
\label{Sec:est}

A typical rotation measurement requires the vector $\bm{\Omega}$ to be imprinted on a (preferably pure) probe state $\ket{\psi}$, in which the latter is shifted by applying a corresponding rotation ${R} (\bm{\Omega}) \in \mathrm{SU}(2)$ that encodes the three parameters $\bm{\Omega}$. A set of measurements is then performed on the output state $\ket{\psi_{\bm{\Omega}}} = {R} (\bm{\Omega}) \ket{\psi}$, with the measurements denoted by a  POVM~\cite{Helstrom:1976ij} $\{ \Pi_{x}\}$, where the POVM elements are labeled by an index $x$ that represents the possible outcomes of the measurement according to Born's rule $p(x |\bm{\Omega}) = \bra{\psi_{\bm{\Omega}}} \Pi_{x} \ket{\psi_{\bm{\Omega}}}$. From here, we infer the vector parameter via an estimator $\widehat{\bm{\Omega}}$~\cite{Kay:1993aa}. 

The performance of the estimator is assessed in terms of the covariance matrix $\mathbf{C}_{\psi} (\widehat{\bm{\Omega}})$, defined as
\begin{equation}
[ \mathbf{C}_{\psi} (\widehat{\bm{\Omega}})]_{jk} = 
\langle ( \widehat{\Omega}_{j} - \Omega_{j}) 
( \widehat{\Omega}_{k} - \Omega_{k} ) \rangle \, ,
\end{equation}
where $j, k \in (1, 2, 3)$ and the expectation value is taken with respect to the probability distribution $p(x|\bm{\Omega})$. The diagonal elements are the variances, the nondiagonal elements characterize the correlations between the estimated parameters, and an ideal estimator will minimize this covariance matrix. 

The ultimate limit for any possible POVM is given by the quantum Cram\'er-Rao bound (QCRB), which promises that~\cite{Braunstein:1994aa}
\begin{equation}
\mathbf{C}_\psi (\widehat{\bm{\Omega}})  \succcurlyeq \mathbf{Q}_{\psi}^{-1}(\bm{\Omega}) \, ,
\end{equation} 
where matrix inequalities $\mathbf{A} \succcurlyeq \mathbf{B}$ mean that $\mathbf{A} - \mathbf{B}$ is a positive semidefinite matrix. Here, the lower bound is the inverse of the quantum Fisher information matrix (QFIM), which takes the particularly simple form for pure states and unitary evolution~\cite{Sidhu:2020aa}
\begin{equation}
\left[\mathbf{Q}_\psi (\bm{\Omega}) \right]_{jk}= 4 \, \mathbf{C}_{\psi} (G_j,G_k) \, .
\end{equation}
The operators $G_{j}$ are the generators of the transformation, determined through $G_{j} =  i R^{\dagger} (\bm{\Omega})\partial_{\Omega_j} R(\bm{\Omega})$, we define the symmetrized covariance between two operators as $\mathbf{C}_{\psi} (A,B)=\frac{1}{2}\langle AB + BA\rangle - \langle A\rangle\langle B\rangle$, and we take expectation values with respect to the original state $\ket{\psi}$.  The quantum Fisher information grows as $\mathbf{Q}\to\nu\mathbf{Q}$ when an experiment is repeated $\nu$ independent times, so we hereafter take $\nu=1$ to inspect the ultimate sensitivity bounds per experimental trial.

Computing these generators requires some subtlety, due to the noncommutativity $[\partial_{\Omega_k} (\mathbf{J}\cdot\mathbf{u}), \mathbf{J}\cdot\mathbf{u}]\neq 0$. Following the approach in Ref.~\cite{Goldberg:2021ab}, one can immediately work out a compact expression for the QFIM:
\begin{equation}
\label{eq:elQFIM}
\mathbf{Q}_{\psi} (\bm{\Omega}) = 4 
\mathbf{H}^{\top} (\bm{\Omega}) 
\, \mathbf{C}_{\psi}(\mathbf{J}) \, \mathbf{H}(\bm{\Omega}) \, ,
\end{equation} 
defining
 $\mathbf{H} (\bm{\Omega})= \begin{pmatrix}\bm{\mathsf{h}}_{\omega} & \bm{\mathsf{h}}_{\Theta}    &  \boldsymbol{\mathsf{h}}_{\Phi} \end{pmatrix}^\top$ with 
\begin{widetext}
\begin{equation}
    \begin{aligned}
    \mathbf{h}_\omega & =-\mathbf{u} \, , \\
    \mathbf{h}_\Theta & =\begin{pmatrix}-\sin \omega \cos \Theta \cos \Phi-\cos \omega \sin \Phi+\sin \Phi  & \qquad (\cos \omega - 1) \cos \Phi-\sin \omega \cos \Theta \sin \Phi & \qquad \sin \omega \sin \Theta\end{pmatrix}^\top \,, \\
    \mathbf{h}_\Phi & =\begin{pmatrix}\sin \omega \sin \Theta \sin \Phi+\sin ^2 \case{1}{2} \omega \sin 2\Theta \cos \Phi & \qquad \sin ^2 \case{1}{2} \omega \sin 2\Theta \sin \Phi-\sin \omega \sin \Theta \cos \Phi & \qquad -2 \sin ^2 \case{1}{2}\omega \sin^2\Theta\end{pmatrix}^\top \, ,
\end{aligned}
\end{equation} 
\end{widetext}
and  $[\mathbf{C}_{\psi} (\mathbf{J}) ]_{jk} = \mathbf{C}_{\psi} (J_{j}, J_{k})$. Notably,
\begin{equation}
\mathbf{H}^{\top} (\bm{\Omega}) 
\,  \mathbf{H}(\bm{\Omega})=\left(
\begin{array}{ccc}
 1 & 0 & 0 \\
 0 & 4\sin^2 \case{1}{2} \omega & 0 \\
 0 & 0 & 4 \sin^2 \case{1}{2} \omega \sin ^2\Theta \\
\end{array}
\right).
\end{equation}

The remarkable property of these expressions is that we have separated the parameter dependence, contained in $\mathbf{H} (\bm{\Omega})$ from the state dependence that is embodied in $ \mathbf{C}_{\psi} (\mathbf{J})$.  To find states optimally suited for estimating arbitrary unknown rotations one must optimize $\mathbf{C}_{\psi} (\mathbf{J})$, which  has been dubbed as the \emph{sensitivity covariance matrix}. The most classical states have the smallest sensitivity covariance matrix, to the point of being singular, while the most quantum states have the largest sensitivity covariance matrix.

Given a covariance matrix, we can balance the precision of the various parameters by using a weight matrix $\mathbf{W} \succ 0$. In this way, the QCRB  leads to the scalar inequality
\begin{equation}
\Tr [ \mathbf{W} \mathbf{C}_{\psi} (\widehat{\bm{\Omega}})] \ge \Tr [ \mathbf{W} \mathbf{Q}_{\psi}^{-1} (\bm{\Omega}) ]  \, . 
\end{equation}
The left-hand side is the so-called weighted mean square error of the estimator, whereas the right-hand side plays the role of a cost function. For a given $\mathbf{W}$, the standard approach is to minimize this cost. Following Ref.~\cite{Goldberg:2021ab}, we take the weight matrix to be the SU(2) metric $\mathbf{W}= \mathbf{H}^\top \mathbf{H}$, so that the QCRB becomes
\begin{equation}
    \Tr [\mathbf{C}_{\psi}^{-1} (\mathbf{J}) ] \geq \frac{9 }{J(J+1)} \, .
    \label{eq:CRB for SU(2)}
\end{equation} 
The trace of the inverse achieves the minimum only when the state is first- and second-order unpolarized; that is, $\langle \mathbf{J} \rangle = 0$ and $\mathbf{C}_{\psi}(\mathbf{J}) \propto \openone $. This is precisely the case for the Kings of Quantumness.

The saturability of the QCRB is a touchy business. For pure states, the QCRB can be saturated if and only if $\IM \langle \psi_{\bm{\Omega}} |L_{j} L_{k} | \psi_{\bm{\Omega}} \rangle = 0$, 
where $L_{j}$ is the symmetric logarithmic derivative respect to the $j$th parameter~\cite{Sidhu:2020aa}.  This hinges upon the expectation values of the commutators of the generators. Fortunately, for states with isotropic covariance matrices, the expectation values of the commutators are guaranteed to vanish. In fact, these expectation values will vanish for all states that are first-order unpolarized. The Kings thus guarantee that all three parameters can be simultaneously estimated at a precision saturating the QCRB for any triad of rotation parameters.

The measurement saturating the QCRB has been recently characterized. However, its experimental implementation may be challenging. Easier is to project the rotated state onto a set of {Bloch coherent states} for various directions and to reconstruct the rotation parameters from these measurements.  

The {set of continuous projections}
\begin{equation}
\label{eq:defHusimi}
    {\mathcal{Q}_{\mathbf{n}}} = | \langle \mathbf{n} |\psi_{\bm{\Omega}} \rangle|^2 
\end{equation} 
constitute the Husimi function~\cite{Husimi:1940aa}. Knowledge of all of the projections $\mathcal{Q}_{\mathbf{n}}$ is equivalent to knowledge of the rotated state $\ket{\psi_{\bm{\Omega}}}$, but such information is redundant: it suffices to sample the function at a few locations ${\mathcal{Q}_{\mathbf{n}}}$ and use these results to orient the Husimi function and thus estimate the rotation parameters. 

At how many locations must the Husimi function be sampled to uniquely orient it? In general, the answer depends on the probe state  and the locations being sampled. Using the same basic principles applied to geographical positioning systems (GPS)~\cite{Hofmann-Wellenhof:2001aa}, we argue that five projections should be more than enough for this orientation problem. 

We first project $\ket{\psi_{\bm{\Omega}}}$ onto an arbitrary Bloch coherent state $\ket{\mathbf{n}_{1}}$, which amounts to sampling the Husimi function at $\mathbf{n}_{1}$. We take this to be the state $|JJ\rangle$. The value of ${\mathcal{Q}_{\mathbf{n}_{1}}}$ defines a set of level curves, and the state \textit{must} be oriented in such a way that $\mathbf{n}_1$ lies on one of these curves. Rotating the state along any of these level curves will produce the same value ${\mathcal{Q}_{\mathbf{n}_{1}}}$.

Next, projecting the rotated state onto another coherent state $\ket{\mathbf{n}_{2}}$ defines another set of level curves. We take this next state to be the opposite coherent state $|J-J\rangle$. Rotating the state along these level curves again produces the same value $\mathcal{Q}_{\mathbf{n}_{2}}$, so in general we expect there to be multiple \textit{intersection points} for orienting the Husimi function such that $\mathbf{n}_{1}$ lies along a curve ${\mathcal{Q}_{\mathbf{n}_{1}}}$ and $\mathbf{n}_{2}$ lies along a curve ${\mathcal{Q}_{\mathbf{n}_{2}}}$.

In all but pathological cases, a third projection uniquely specifies one of the above intersection points for orienting the Husimi function. We include a fourth projection to  deal with pathological cases and a fifth projection to help with normalization. We take these projections to be onto three coherent states pointing toward the equator, first in the $y$-direction, then in the $x$-direction, and finally in the $\sqrt{2}x+y$-direction.
Put in different words,  the set of $d$ angular coordinates $\mathbf{n}_i$ can be rigidly rotated until the $d$ projections ${\mathcal{Q}_{\mathbf{n}_{i}}}$ match the given state $\ket{\psi_{\bm{\Omega}}}$.  The pathological cases are those for which the projections match the given state rotated by different sets of rotation parameters. 


\section{Experiment}
\label{Sec:exp}

\begin{figure}
    \centering
    \includegraphics[width=.85\columnwidth]{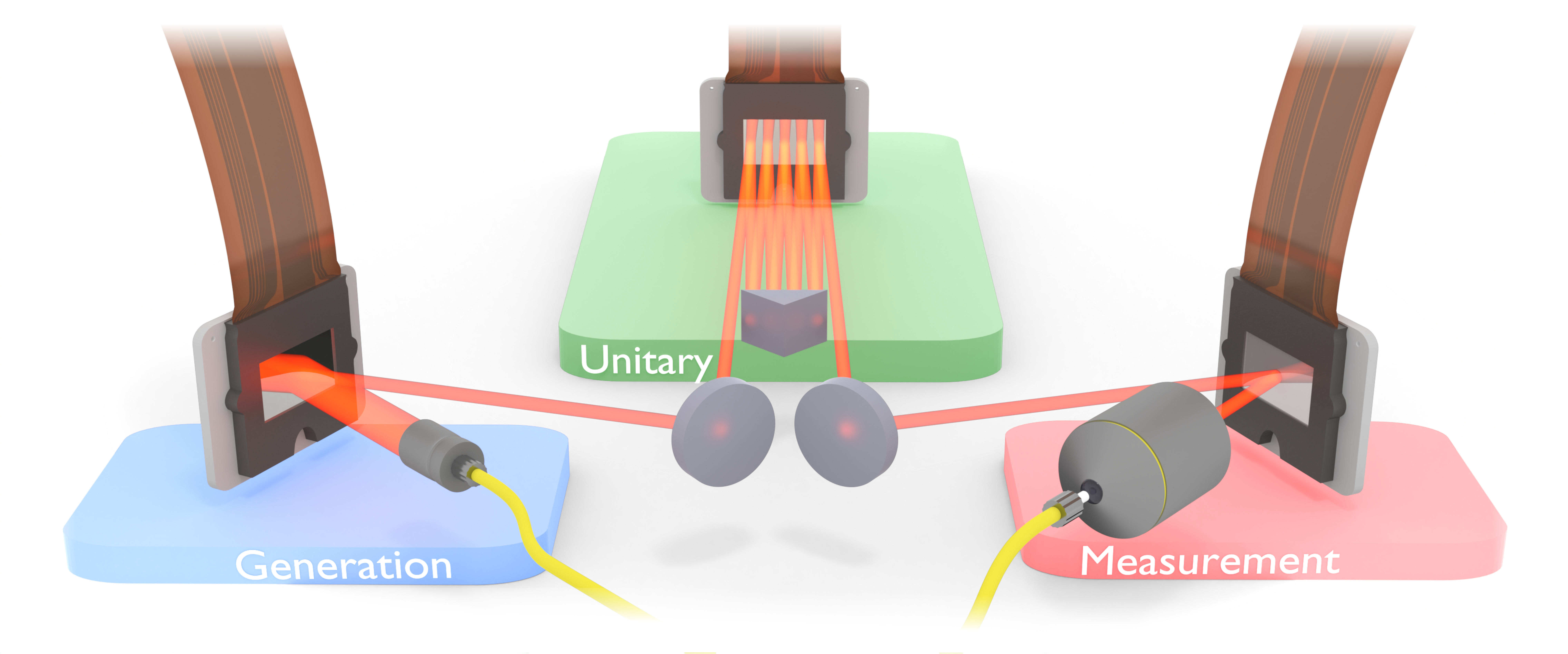}
    \caption{A simplified sketch of the experimental system. Three SLM screens are used to perform the three steps of the experiment. The first one transforms the Gaussian beam coming out of a single mode fiber (SMF, yellow) into the desired King state that is then rotated using a unitary operation implemented through MPLC on the second SLM. The third SLM performs a projective measurement onto a state with the aid of an SMF.}
    \label{fig:setup}
\end{figure}

To verify the proposed method, we use an experimental setup sketched in Fig.~\ref{fig:setup}. It consists of three sections: state generation, unitary manipulation, and state measurement.
Different $4f$ imaging systems, which were omitted from the figure, are used to connect the sections.  We use a CW diode laser (Roithner RLT808-100G, $\Delta \lambda = 2$~nm) at a central wavelength of 810~nm, and phase-only spatial light modulators (SLM, Holoeye Pluto-2) along with standard free-space optical components to perform the experiment. For accurate phase modulations, we also perform aberration correction for all phase screens using a Gerchberg-Saxton phase retrieval algorithm~\cite{Jesacher:2007aa}.

First, we encode the probe state $\ket{\psi}$ in the transverse spatial degree of freedom of a laser beam; i.e., Laguerre-Gauss modes carrying OAM of light, emerging from a single-mode fiber by displaying a phase and amplitude modulating mask on the first SLM~\cite{Bolduc:2013aa} with an added Gaussian correction~\cite{Plachta:2022aa,Hiekkamaki:2022aa}.
The chosen states are Kings of Quantumness: we consider here the $J=2$ state with tetrahedral symmetry that lives in a 5-dimensional Hilbert space
\begin{equation}
|\psi\rangle=\frac{\sqrt{2}|2\, -\!1\rangle+|2\,2\rangle}{\sqrt{3}}
\end{equation} 
and the $J=3$ state formed from a square-based pyramid and its reflection that lives in a 7-dimensional Hilbert space
\begin{equation}
|\psi\rangle=\frac{|3\, -\!2\rangle-|3\,2\rangle}{\sqrt{2}}.
\end{equation} 
This encoded state is then imaged onto the first phase screen of an MPLC system, which consists of 5 consecutive phase modulations implemented using a single SLM screen~\cite{Labroille:2014aa}. The MPLC system is capable of performing arbitrary unitary transformations in the transverse spatial degree of freedom~\cite{Brandt:2020aa}, using phase modulations calculated through a wavefront-matching algorithm \cite{Fontaine:2019aa}. Here, the MPLC is used to realize the unitary transformations $R(\bm{\Omega})$ of the probe state $R(\bm{\Omega})\ket{\psi} = \ket{\psi_{\bm{\Omega}}}$, i.e., rotations of the state. After the MPLC, the rotated state is imaged onto the measurement SLM, which is used along with an SMF to perform projective measurements of the rotated probe state $\mathcal{Q}_{\mathbf{n}} = | \langle \mathbf{n} |\psi_{\bm{\Omega}} \rangle|^2 $.

The raw data of the projective measurements consists of power values with different measurement settings.  In the ideal case, the projective measurement is simply given by $\mathcal{Q}_{\mathbf{n}} = | \langle \mathbf{n} |\psi_{\bm{\Omega}} \rangle|^2  = P_{\langle \mathbf{n} |\psi_{\bm{\Omega}} \rangle}/P_{\langle \psi_{\bm{\Omega}} |\psi_{\bm{\Omega}} \rangle}$, where $P_{\langle \mathbf{n} |\psi_{\bm{\Omega}} \rangle}$ is the power coupled to the SMF when projecting the rotated probe state $\ket{\psi{_{\bm{\Omega}}}}$ onto the Bloch coherent state $\ket{\mathbf{n}}$, and $P_{\langle \psi_{\bm{\Omega}} |\psi_{\bm{\Omega}} \rangle}$ is the total power readout when projecting the rotated probe state onto itself, taking into account the efficiency of the measurement.

However, in our projective measurement scheme, the power that is coupled to the SMF is highly dependent on the state being generated and projected on. To account for this, we must measure and compensate for these state-dependent detection efficiencies $\eta$. To measure the detection efficiencies of the Bloch coherent states, we make use of the following scheme. 
We generate the Bloch coherent state with the first SLM, and image the state through the system unaltered (setting MPLC to perform an identity unitary).  Then, we measure the power projected onto the Bloch coherent state itself $P_{\langle \mathbf{n} |\mathbf{n} \rangle}$, and the total power of the beam before the final SLM $P_{\mathbf{n}}$, giving us an efficiency measure $\eta_{\mathbf{n}} = P_{\langle \mathbf{n} |\mathbf{n} \rangle}/P_{\mathbf{n}}$. For the detection efficiencies of the rotated probe states $\eta_{\psi_{\bm{\Omega}}}$, we make use of a similar scheme, but instead of imaging the state through the system unaltered, we generate a probe state $\ket{\psi}$ with the first SLM, rotate the state using the MPLC $R(\bm{\Omega})\ket{\psi} = \ket{\psi_{\bm{\Omega}}}$, and measure the detection efficiency  $\eta_{\psi_{\bm{\Omega}}} = P_{\langle \psi_{\bm{\Omega}} |\psi_{\bm{\Omega}} \rangle}/P_{\psi_{\bm{\Omega}}}$. With these detection efficiencies, the projective measurements are given by $\mathcal{Q}_{\mathbf{n}} = \eta_{\psi_{\bm{\Omega}}}P_{\langle \mathbf{n} |\psi_{\bm{\Omega}} \rangle}/(\eta_{\mathbf{n}}P_{\langle \psi_{\bm{\Omega}} |\psi_{\bm{\Omega}} \rangle})=P_{\langle \mathbf{n} |\psi_{\bm{\Omega}} \rangle}/(\eta_{\mathbf{n}}P_{\psi_{\bm{\Omega}}})$; i.e., the fraction of the power 
coupled to the SMF, scaled by the efficiency of projecting onto the coherent state in question. 
\begin{figure*}
    \centering
    \includegraphics[width=1.85\columnwidth]{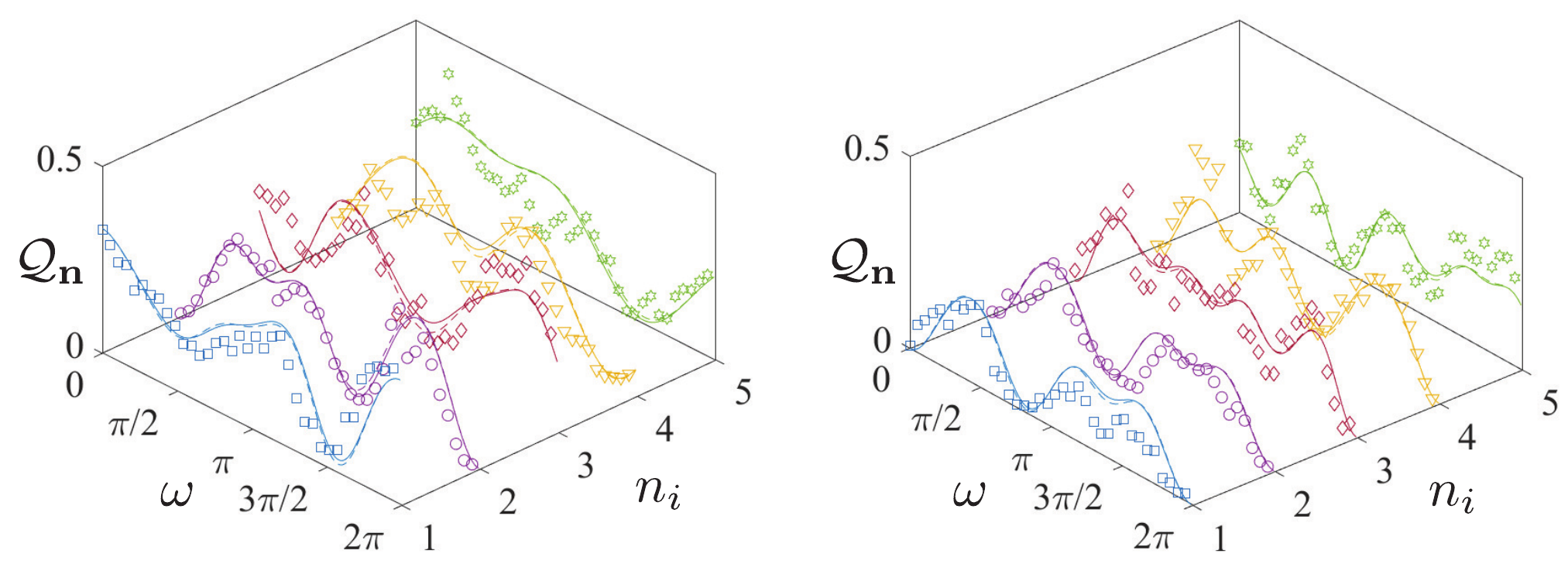}
    \caption{Projections of the rotated $J=2$ (left) and $J=3$ (right) Kings of Quantumness onto five different spin coherent states. The plotted error bars are 10 times the measured standard deviations; however, in general, they are too tiny to be visible. The dashed line shows the estimated curve, while the full line shows the theoretically expected results.}
    \label{fig:5D all projections with 10x errors}
\end{figure*}
Each of the power values is measured as a sample mean of power data gathered for half a second, corresponding to approximately 50 datapoints. From these data, we also calculate the standard deviation for each power value, which are used to infer the standard deviations of the projective measurements via error propagation. The projective measurement data for rotated $J=2$ and $J=3$ Kings of Quantumness, along with their standard deviations, are presented in Fig.~\ref{fig:5D all projections with 10x errors}. However, even after tenfold multiplication of the error bars, they are too small to be visible.

\section{Results}
\label{Sec:res}

\subsection{Axis estimation}
Figure~\ref{fig:5D all projections with 10x errors} shows the data for the $J=2$ King of Quantumness state, i.e. the tetrahedron initial state, and projections onto five coherent states for the same randomly chosen axis variables $(\Theta,\Phi)=(1.11,3.75)$ and all rotation angles in intervals of $\pi/18$, i.e., 10 degrees.  Combining all of these data for all 37 of the rotation angles, we can perform maximum likelihood estimation following standard procedures~\cite{Hradil:2006aa,Rehacek:2008aa}
\begin{equation}
(\hat{\Theta},\hat{\Phi})=\arg\max \sum_{i}\sum_{j=1}^5 \mathcal{Q}_{\mathbf{n}_j}\log\frac{|\braket{\mathbf{n}_j|R(\omega_i,\mathbf{u})|\psi}|^2}{\sum_{i^\prime,j^\prime}|\braket{\mathbf{n}_{j^\prime}|R(\omega_{i^\prime},\mathbf{u})|\psi}|^2}.
\label{eq:axis MLE argmax}
\end{equation} 
With \textit{no further constraints} on $0\leq \Theta\leq \pi$ and $0\leq \Phi<2\pi$, the maximization procedure finds $(\hat{\Theta},\hat{\Phi})=(1.13,3.79)$, immediately demonstrating the usefulness of this procedure. We note that there is a local maximum at the true variables (1.11,3.75) but this is not the global maximum.  To further demonstrate the very good agreement, we also plot the estimated lines versus the true lines for coherent-state projections of the state rotated about the estimated versus the true axis in Fig. \ref{fig:5D all projections with 10x errors}. 

We can consider the Fisher information matrix of the estimated parameters with components
\begin{equation}
    {F}_{ij}\approx \sum_{kl} \frac{\mathcal{Q}}{\mathcal{Q}_{\mathbf{n}_l}}\frac{\partial}{\partial \Omega_i}\frac{p_{kl}}{P}\frac{\partial}{\partial \Omega_j}\frac{p_{kl}}{P},
\end{equation}  
corresponding to the amount of information per probe state and estimate the uncertainty from the inverse of the Fisher information matrix.
Here,
\begin{equation}
p_{kl}=|\braket{\mathbf{n}_l|R(\omega_k,\mathbf{u})|\psi}|^2, \quad
P=\sum_{kl} p_{kl} \, , \quad \mathcal{Q} =\sum_{l} \mathcal{Q}_{\mathbf{n}_l}\, ,
\end{equation} 
and all of the derivatives can be computed analytically using the generators $G_j=\mathbf{J}\cdot\mathbf{h}_j$. When we do this, we find
\begin{equation}
 \mathbf{F}^{-1} (\Theta,\Phi) = \left(
\begin{array}{cc}
 0.261 & -0.0471 \\
 -0.0471 & 0.452 \\
\end{array}
\right).
\end{equation}  
Using this inverse as the covariance matrix for the estimates of the axis parameters~\cite{Rehacek:2008aa} lets us report the uncertainties in our estimates of $\Theta$ and $\Phi$ as $0.51$ and $0.67$, respectively.
For comparison, the best possible value of the QFI for a single experiment is $\mathsf{\mathbf{F}}_\mathrm{max}= \case{1}{3} 4J(J+1)\sin^2 (\omega_i/2) \mathrm{diag}(4,4\sin^2\Theta)$, where $\mathrm{diag} (\mathbf{v})$  is a square diagonal matrix with the elements of vector $\mathbf{v}$ on the main diagonal. Averaging over the 37 different values of $\omega$ that were used, this becomes 
\begin{equation}
\mathbf{F}_{\mathrm{max}}^{-1}(1.13,3.79) = \left(
\begin{array}{cc}
 0.0642 & 0. \\
 0. & 0.0786 \\
\end{array}
\right).
\end{equation}  
Our measured uncertainties are larger than the ultimate bounds by factors of 2.0 and 2.4 for $\Theta$ and $\Phi$, respectively.
 
In the second set of experiments, we performed the same tasks for the $J=3$ state. The true rotation angles are $(\Theta,\Phi)=(2.40, 2.76)$ and the estimated ones with no constraints are $(\Theta,\Phi)=(2.37, 2.75)$. Again, the data along with the estimated and theoretical curves are plotted in Fig.~\ref{fig:5D all projections with 10x errors}, showing good agreement.  However, due to the increased dimension of the utilized states without an increase in phase modulations planes of the MPLC system, the performance of our experiment slightly degrades.
Performing the observed Fisher information calculation yields
\begin{equation}
\mathbf{F}^{-1}(\Theta,\Phi) =\left(
\begin{array}{cc}
 0.203 & 0.0174 \\
 0.0174 &  0.282 \\
\end{array}
\right),
\end{equation} 
to be compared to the theoretical minimum uncertainty given by
\begin{equation}
\mathbf{F}_\mathrm{max}^{-1}(2.37, 2.75) =\left(
\begin{array}{cc}
 0.0321 & 0 \\
 0 & 0.0666 \\
\end{array}
\right).
\end{equation}  
Our measured uncertainties are larger than the ultimate bounds by factors of 2.5 and 2.1 for $\Theta$ and $\Phi$, respectively.

\begin{figure*}
    \centering
    \includegraphics[width=1.85\columnwidth]{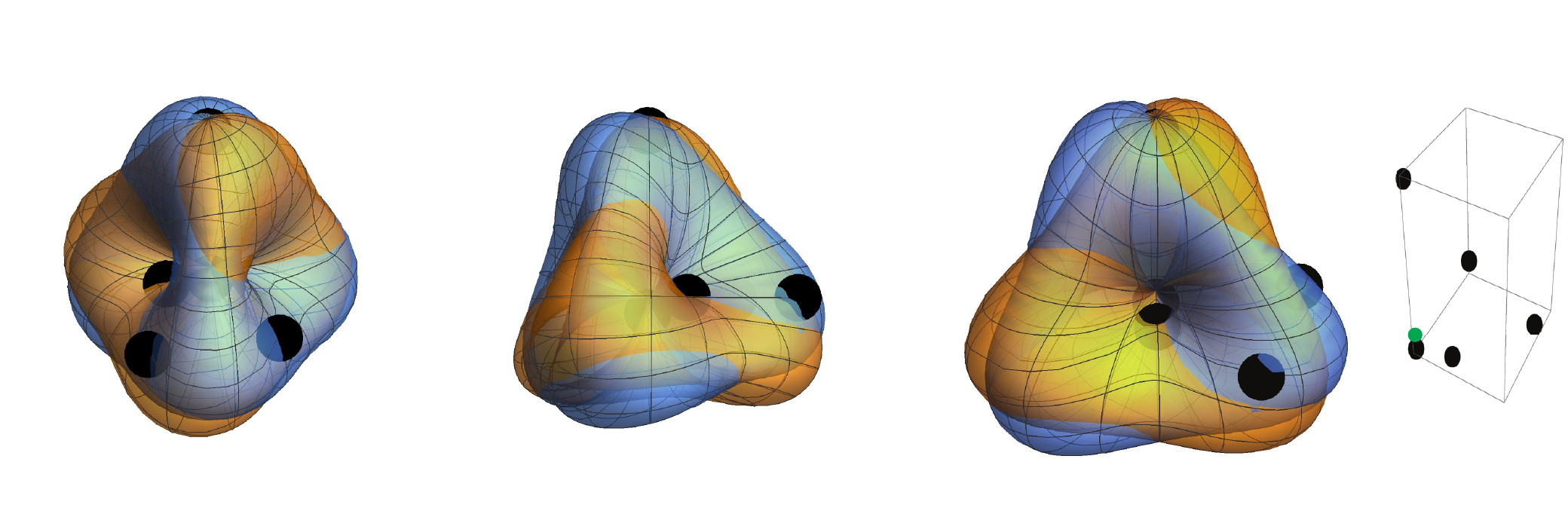}
    \caption{Example of the true rotated state (orange surface) and estimated rotated state (blue surface) for $J=2$ and a rotation by $\pi/18$ about a randomly chosen rotation axis. The radial coordinates are given by the magnitude of the Husimi $\mathcal{Q}$-function. Plotted are three perspectives of this three-dimensional object, along with five black spheres at the five measured data points. The rotation estimation algorithm tries to rotate the surface until it best matches the data points, with ``best'' defined according to a maximum likelihood estimation procedure.  Plotted on the right is a three-dimensional box to show how the black spheres are distributed in space, along with a small green sphere at the origin toward the bottom-left corner. }
    \label{fig:rot depiction t10}
\end{figure*}
\begin{figure}[b]
    \centering
    \includegraphics[width=.85\columnwidth]{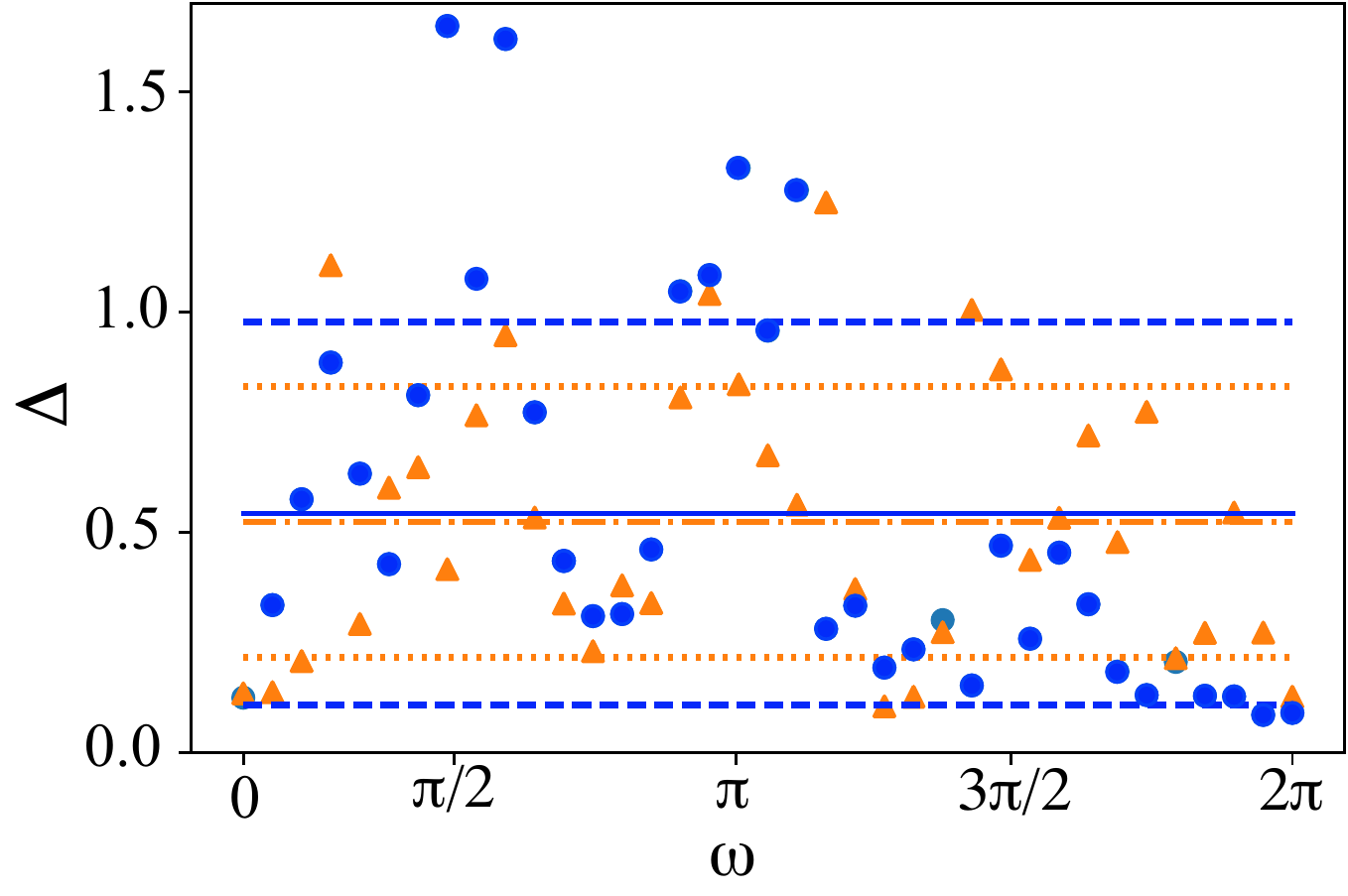}
    \caption{Angular deviation between the true rotation and estimated rotation when all three parameters of the rotation are simultaneously estimated. The probe states are a King state with $J=2$ (i.e., a state with tetrahedral symmetry; blue circles) and $J=3$ (i.e., an octahedral state with $Z_2\times S_4$ symmetry; orange triangles). There are 37 different experiments, each with the same to-be-determined axis of rotation and all rotation angles equally spaced from 0 to $2\pi$. The mean is plotted as the solid (dot-dashed) line and the one standard deviation away as the dashed (dotted) lines for $J=2$ ($J=3$). The small errors, especially for rotation angles close to 0 and $2\pi$, validate this method for estimating all three parameters of a rotation.}
    \label{fig:rot errors N4 N6}
\end{figure}

We can discuss the scaling of these results with $J$. While there is clearly a larger Fisher information for the $J=3$ state than the $J=2$ state, we must be careful because the best possible QFI depends on the rotation axis. We can create a unique figure of merit by weighing the observed uncertainties by the metric $\mathfrak{g}=\mathrm{diag}(1,\sin^2\Theta)$ and taking the trace. This finds
\begin{equation}
    \Tr(\mathfrak{g} \mathbf{F}_{J=2}^{-1})=0.63 \, ,
    \qquad
    \Tr(\mathfrak{g}  \mathbf{F}_{J=3}^{-1})=0.34.
\end{equation}  
This decrease in total uncertainty with $N$ is approximately the anticipated $1/[J(J+1)]$ scaling, which would provide a factor of 2 difference here. For comparison, the ultimate limits are $37/[48 J(J+1)]$, which approximately equal 0.13 and 0.064 for $J=2$ and $J=3$. A NOON state, with $\mathbf{C}_\psi=\mathrm{diag}(J/2,J/2,J^2)$ or a transposition of the diagonal elements for a NOON state in another direction, has ultimate limits in the range $0.18-0.22$ and $0.10-0.14$ for $J=2$ and $J=3$, respectively, for the axes chosen randomly here.
Since the results remain within a factor of 3 in uncertainty of the best possible values for all $J$ tested here, it is certainly reasonable to expect dramatic quantum advantages as $J$ grows, at which point they will also outperform NOON states. All of these outperform SU(2)-coherent states, the most classical of states, because these cannot be used to simultaneously estimate multiple parameters of a rotation due to the singularity of their sensitivity covariance matrices.

\subsection{Angle and axis estimation}
Next, we can do this estimation $37$ times to find all three rotation parameters for each specific rotation. Such an estimation cannot be done in a single trial using (classical) SU(2)-coherent states, as those are insensitive to one of the three parameters of a rotation. The maximization is the same as in Eq. \eqref{eq:axis MLE argmax}
but removing the sums over $i$ and $i^\prime$ and repeating the optimization for each value of $i=i^\prime$:
\begin{equation}
\widehat{\bm{\Omega}}_i=\arg\max \sum_{j=1}^5 \mathcal{Q}_{\mathbf{n}_j}\log\frac{|\braket{\mathbf{n}_j|R(\bm{\Omega}_i)|\psi}|^2}{\sum_{j^\prime}|\braket{\mathbf{n}_{j^\prime}|R(\bm{\Omega}_i)|\psi}|^2}.
\end{equation}

Figure~\ref{fig:rot depiction t10} depicts the optimization process geometrically for the first nonzero rotation with $J=2$ for a rotation by $\pi/18$ ($10^\circ$). The true rotated state has a particular $\mathcal{Q}$-function that is sampled at five points, then the optimization algorithm rotates the original state until it best aligns with those five sampled points. The offset between the true and estimated state as well as how closely each matches the level curves from the five data points can be inspected visually. Throughout, one can see that the $Q$-function varies dramatically with angular coordinates, possessing tetrahedral symmetry, which is what makes this state so useful for estimating rotations.

It is cumbersome to visualize all 37 experiments for each of the tested values of $J$, so we require a method to aggregate the estimation results. We choose to investigate the deviation between the true rotation and the estimated rotation, quantified by the angle of rotation one would require to convert between the true and estimated rotation. This can be given by the unitary rotation operator
\begin{equation}
R(\Delta_i,\mathbf{u}_{i})\equiv R(\bm{\Omega}_i)^\dagger R(\widehat{\bm{\Omega}}_i).
\end{equation} 
We need not worry about the direction $\mathbf{u}_i$ in characterizing this error. To find $\Delta_i$, we use the Hilbert-Schmidt norm $\Tr[ R(\bm{\Omega}_i)^\dagger R(\widehat{\bm{\Omega}}_i)]=\Tr[R(\Delta_i,\mathbf{u}_{i})]=\cos\Delta_i J+\sin\Delta_i J/\tan(\Delta_i/2)$, found using invariance of the trace under unitary transformations such that the trace is evaluated in the basis of eigenstates of $\mathbf{J}\cdot\mathbf{u}_i$, and solve for the smallest value of $\Delta_i$. In doing so, we find the average deviations $\Delta=0.54\pm0.43$ and $\Delta=0.52\pm0.31$ for $J=2$ and $J=3$, respectively, with all of the values plotted in Fig.~\ref{fig:rot errors N4 N6}. These are all much smaller than might be found for a random rotation, where a random rotation would have deviation $\pi/2$ on average, showcasing the usefulness of our method.

\section{Concluding remarks}
\label{Sec:concl}

 Rotations constitute the epitome of how quantum properties can boost sensitivity and precision. We have introduced a simple measurement scheme that achieves results close to the ultimate bounds dictated by quantum theory, without requiring complicated entangled measurement strategies, and for simultaneously estimating all three parameters of a rotation. We expect these results to be relevant in  years to come.

\acknowledgements{ME acknowledges support from the Academy of Finland through the project BIQOS (Decision 336375). AZG and FB acknowledge that the NRC headquarters is located on the traditional unceded territory of the Algonquin Anishinaabe and Mohawk people and support from NRC's Quantum Sensors Challenge Program. AZG acknowledges funding from the NSERC PDF program. MH acknowledges the Doctoral School of Tampere University and the Magnus Ehrnrooth foundation. LLSS acknowledges support from Ministerio de Ciencia e Innovaci\'on (Grant  PID2021-127781NB-I00). RF acknowledges support from the Academy of Finland through the Academy Research Fellowship (Decision 332399). ME, MH, RF acknowledge the support of the Academy of Finland through the Photonics Research and Innovation Flagship (PREIN - decision 320165).}


%

\end{document}